\documentstyle[editedvolume]{crckapb}

\makeatletter                                            
\@ifundefined{epsfbox}{\@input{epsf.sty}}{\relax}        
\def\plotone#1{\centering \leavevmode                    
\epsfxsize=\columnwidth \epsfbox{#1}}                    
\def\plotone_reduction#1#2{\centering \leavevmode        
\epsfxsize=#2\columnwidth \epsfbox{#1}}                  
\makeatother                                             

\begin{opening}
\title{Light Propagation in Inhomogeneous Universe \protect\\}
\subtitle{Preliminary Results from Gravitational Lensing Experiments}
\author{Premana W. Premadi}
\institute{Institute of Astronomy, Tohoku University\\
           Sendai 981, Japan}
\author{Hugo Martel}
\institute{Dept. of Astronomy, University of Texas\\
	    Austin, TX 78712, USA}
\author{Richard A. Matzner}
\institute{Center for Relativity, University of Texas\\
	    Austin, TX 78712, USA}
\end{opening}

\runningtitle{Light Propagation in Inhomogeneous Universes}

\begin{document}

\section{Overview}

Using a multi-plane lensing method that we have
developed (Premadi, Martel, \& Matzner 1998), we follow the evolution
of light beams as they propagate through
inhomogeneous universes. We use a P$^3$M 
code to simulate the formation and evolution of large-scale structure.
The resolution of the simulations is increased to
sub-Megaparsec scales by using a Monte Carlo method to locate galaxies
inside the computational volume according to the underlying particle
distribution.

We consider cold dark matter models normalized to {\it COBE}, 
and perform a large parameter survey by varying the cosmological parameters
$\Omega_0$, $\lambda_0$, $H_0$, and $n$ (the tilt of the primordial power
spectrum). This parameter survey is still in progress.
Table 1 gives the values of the
parameters for the models we have studied so far.

\begin{table}[htb]
\begin{center}
\caption{The Cosmological Parameters}
\begin{tabular}{llllllllll}
\hline
$\Omega_0$ & $\lambda_0$ & $H_0$ & $n$ & $\sigma_8\qquad$ &
$\Omega_0$ & $\lambda_0$ & $H_0$ & $n$ & $\sigma_8$ \\
\hline
1.0 & 0.0 & 65.0 & 0.7698 & 1.0 & 0.2 & 0.0 & 65.0 & 1.3188 & 0.5 \\
1.0 & 0.0 & 75.0 & 0.7094 & 1.0 & 0.2 & 0.0 & 75.0 & 1.2190 & 0.5 \\
1.0 & 0.0 & 65.0 & 0.8506 & 1.2 & 0.5 & 0.5 & 65.0 & 0.7808 & 0.8 \\
1.0 & 0.0 & 75.0 & 0.7893 & 1.2 & 0.5 & 0.5 & 75.0 & 0.7049 & 0.8 \\
0.5 & 0.0 & 65.0 & 0.9457 & 0.8 & 0.5 & 0.5 & 65.0 & 0.8807 & 1.0 \\
0.5 & 0.0 & 75.0 & 0.8686 & 0.8 & 0.5 & 0.5 & 75.0 & 0.8024 & 1.0 \\
0.5 & 0.0 & 65.0 & 1.0439 & 1.0 & 0.2 & 0.8 & 65.0 & 0.9326 & 0.6 \\
0.5 & 0.0 & 75.0 & 0.9656 & 1.0 & 0.2 & 0.8 & 75.0 & 0.8273 & 0.6 \\
0.2 & 0.0 & 65.0 & 1.0966 & 0.3 & 0.2 & 0.8 & 65.0 & 1.0702 & 0.8 \\
0.2 & 0.0 & 75.0 & 0.9993 & 0.3 & 0.2 & 0.8 & 75.0 & 0.9629 & 0.8 \\
\hline
\end{tabular}
\end{center}
\end{table}

\begin{figure}
\epsfxsize=11cm
\vspace{-1.9cm}
\hspace{0.6cm}\epsfbox{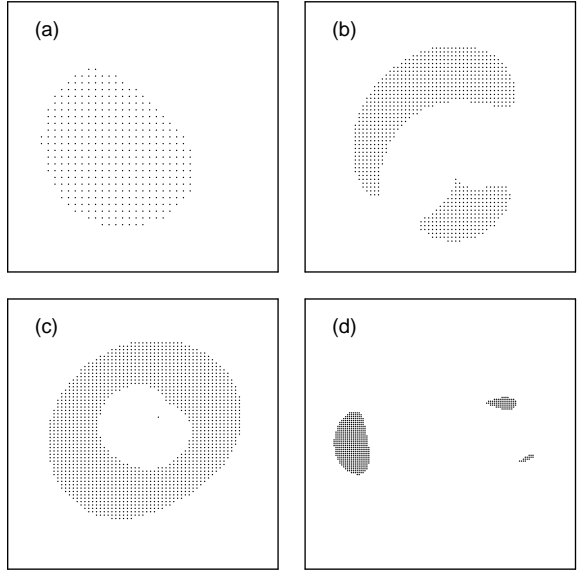}
\vspace{-5.1cm}
\caption[h]{Images of distant circular sources}
\end{figure}

\section{The Ray Shooting Experiments}

For each model, we perform numerous ray-tracing experiments, 
propagating beams of $31^2$ light rays back in time
up to redshifts $z=3$ or 5. From the results of these
experiments, we build statistics of
the shear and magnification of sources, as well as the properties
of images. In Figure 1, we plot the images of distant
circular sources, to illustrate interesting cases that occurred in some
experiments: magnification and shear (Fig. 1a), double image (Fig. 1b),
Einstein ring (Fig. 1c), and triple image (Fig. 1d).

\end{document}